\documentclass[12pt]{article}
\usepackage{amsfonts}
\usepackage{graphicx}
\usepackage{amsthm}
\usepackage{amsmath}
\usepackage{fancyhdr}
\usepackage[left = 35mm,right=25mm,top=25mm]{geometry}
\usepackage{rotating}
\usepackage{pstricks}
\usepackage[agsm]{harvard}
\begin{document}

\newcommand{\ctau}{\hbox{\it$\tau$}}
\newcommand{\cov}{\mbox{cov}}
\newcommand{\diag}{\mbox{diag}}
\newcommand{\Var}{\mbox{Var}}
\newcommand{\E}{\mbox{E}}
\newcommand{\Cov}{\mbox{Cov}}
\newcommand{\Corr}{\mbox{Corr}}
\newcommand{\diff}{\mbox{d}}
\newcommand{\otherwise}{\mbox{otherwise}}
\newcommand{\boldtheta}{\mbox{\boldmath{$\theta$}}}
\newcommand{\xstarboldtheta}{\mathbf{x} = \mathbf{x}^*(\theta)}
\newcommand{\boldlambda}{\mbox{\boldmath{$\lambda$}}}
\newcommand{\boldLambda}{\mbox{\boldmath{$\Lambda$}}}
\newcommand{\boldbeta}{\mbox{\boldmath{$\beta$}}}
\newcommand{\boldmu}{\mbox{\boldmath{$\mu$}}}
\newcommand{\boldvarepsilon}{\mbox{\boldmath{$\varepsilon$}}}
\newcommand{\boldOmega}{\mbox{\boldmath{$\Omega$}}}
\newcommand{\boldomega}{\mbox{\boldmath{$\omega$}}}
\newcommand{\boldgamma}{\mbox{\boldmath{$\gamma$}}}
\newcommand{\boldxi}{\mbox{\boldmath{$\xi$}}}
\newcommand{\boldSigma}{\mbox{\boldmath{$\Sigma$}}}
\newcommand{\boldrho}{\mbox{\boldmath{$\rho$}}}
\newcommand{\boldphi}{\mbox{\boldmath{$\phi$}}}
\newcommand{\boldpi}{\mbox{\boldmath{$\pi$}}}
\newcommand{\mle}{\mbox{m.l.e.}}
\newcommand{\iid}{\mbox{i.i.d.}}
\newcommand{\Like}{\mathcal{L}}
\newcommand{\normal}{\mbox{N}}
\newcommand{\by}{\mathbf{y}}
\newcommand{\be}{\mathbf{e}}
\newcommand{\bw}{\mathbf{w}}
\newcommand{\bu}{\mathbf{u}}
\newcommand{\bX}{\mathbf{X}}
\newcommand{\bx}{\mathbf{x}}
\newcommand{\bV}{\mathbf{V}}
\newcommand{\bz}{\mathbf{z}}
\newcommand{\transpose}{\mbox{\tiny T}}
\newcommand{\VO}{$\dot{\mbox{V}}\mbox{O}_2 $ \hspace{0.1mm}}
\newcommand{\VCO}{$\dot{\mbox{V}}\mbox{CO}_2$ \hspace{0.1mm}}
\theoremstyle{plain}
\newtheorem{result}{Result}

\title{Approximate simulation-free Bayesian inference for multiple changepoint models with dependence within segments}
\author{Jason Wyse$^1$, Nial Friel$^2$ and H{\aa}vard Rue$^3$
\\
\small $^1$University College London, London, UK\\
\small $^2$University College Dublin, Belfield, Dublin 4, Ireland\\
\small $^3$ Norwegian University of Science and Technology, Trondheim, Norway}
\date{May 2011}

\maketitle
\normalsize

\begin{abstract}
This paper proposes approaches for the analysis of multiple changepoint models when dependency in the data is modelled through a hierarchical Gaussian Markov random field. Integrated nested Laplace approximations are used to approximate data quantities, and an approximate filtering recursions approach is proposed for savings in compuational cost when detecting changepoints. All of these methods are simulation free. Analysis of real data demonstrates the usefulness of the approach in general. The new models which allow for data dependence are compared with conventional models where data within segments is assumed independent.
\end{abstract}

\section{Introduction}

There is a substantial volume of literature devoted to the estimation of multiple changepoint models. These models are used frequently in econometrics, signal processing and bioinformatics as well as other areas. The idea is that ``time'' ordered data (where time may be fictitious and only refers to some natural ordering of the data) is assumed to follow a statistical model which undergoes abrupt changes at some time points, termed the changepoints. The changepoints split the data into contiguous segments. The parametric model assumed for the data usually remains the same accross segments, but changes occur in its specification. For example, in the famous coal mining disasters data~\cite{Jarrett79}, disasters are usually assumed to follow a Poisson distribution where the rate of this distribution undergoes abrupt changes at specific timepoints.~\citeasnoun{Fearnhead06} discusses how to perform exact simulation from the posterior distribution of multiple changepoints for a specific class of models using recursive techniques based on filtering distributions. The class of models considered assumes data is independent within a homogeneous segment and the prior taken on the unknown model parameters for that segment allows analytical evaluation of the marginal likelihood for that segment. The paper of~\citeasnoun{Fearnhead06} proposes a very promising step forward for the analysis of multiple changepoint models, where the number of changepoints is not known beforehand. The methods developed there allow for efficient simulation of large samples of changepoints without resorting to MCMC. 

An obstacle which may prevent wide applicability of the methods discussed in~\citeasnoun{Fearnhead06}, is the requirement that the assumed model must have a segment marginal likelihood which is analytically tractable. However, such a requirement can usually not be fulfilled by models which allow for data dependency within a segment, a desirable assumption in many situations. Dependency is possible across regimes in some cases (see~\citeasnoun{Fearnhead10}), but the assumption of independent data still holds. The main aim of this paper is to provide a solution to these issues and open up the opportunity for more complex segment models which allow for temporal dependency between data points. This is achieved by hybridizing the methods in~\citeasnoun{Fearnhead06} and recent methodology for the approximation of Gaussian Markov random field (GMRF) model quantities due to~\citeasnoun{Rue09} termed INLAs (integrated nested Laplace approximations).

The INLA methodology provides computationally efficient approximations to GMRF posteriors, which have been demonstrated to outperform MCMC in certain situations~\cite{Rue09}. An 
advantage to such approximations is that they avoid lengthly MCMC runs to fully explore the posterior support and they also avoid the need to demonstrate that these runs have converged. Another advantage is that the approximations may be used to estimate quantities such as the marginal likelihood of the data under a given GMRF model, the quantity which is of main interest here to overcome the requirement of an analytically tractable segment marginal likelihood in~\citeasnoun{Fearnhead06}.

The R-INLA package~\citeasnoun{Rue09} for R-2.11.1 may be used to do all of the aforementioned approximations for a range of GMRF hierarchical models. It aims to give an off-the-shelf tool for INLAs. Currently the package implements many exponential family models; Gaussian with identity-link; Poisson with log-link; Binomial with logit-link; for many different temporal GMRFs; random effects models; first order auto-regressive; first and second order random walk (neither of these lists are exhaustive). The package also implements spatial GMRFs in two and three dimensions and is currently still evolving with new additions on a regular basis. Use of this package avoids programming for specific models as it allows the selection of any observational data model and selection of the desired GMRF through a one line call to the R-INLA package. The R-INLA package is used for all the computations on hierarchical GMRF models in this paper. 
 
The remainder of this paper is organised as follows. 
 Section~\ref{sec:INLA_CP_changepoint_models} gives a brief review of recursions for performing inference conditional on a particular number of changepoints as given in~\citeasnoun{Fearnhead06}. In Section~\ref{sec:INLA_CP_RFRs} possible computational difficulties are discussed and solutions for these are proposed. Sections~\ref{sec:INLA_CP_coal_mining} and~\ref{sec:INLA_CP_well_log_data} analyze real data examples; the coal-mining data is analyzed using a model with dependency and this is compared with the analysis of~\citeasnoun{Fearnhead06}; and Well-log data~\cite{ORuanaidh96} is analyzed with a model that allows for dependency between adjacent data points, such that the dependency relation may change across segments.  Section~\ref{sec:INLA_CP_stochastic_volatility} explores the possibility of detecting changepoints under the assumption of a stochastic volatility model. The paper concludes with a discussion.

\section{Changepoint models} \label{sec:INLA_CP_changepoint_models}

~\citeasnoun{Fearnhead06} gives a detailed account of how filtering recursions approaches may be applied in changepoint problems. Some of the models considered there used a Markov point process prior for the number and position of the changepoints. ~\citeasnoun{Wyse10} demonstrated that the posterior distribution may sometimes be sensitive to the choice of the parameters for the point process. In this paper, the focus will be on performing inference for the changepoint positions after estimating the most probable number of changepoints {\it a posteriori}, although it is noted that the methods also apply to the case of a point process prior.  Denote $k$ ordered changepoints by $\tau_1,\dots,\tau_k$. The likelihood for the data $\by_{1:n}$, conditional on the $k$ changepoints and the latent field $\bx$, assuming segments are independent of one another is
\[
\pi(\by|\bx,\boldtheta) = \prod_{j=1}^{k+1} \pi(\by_{\tau_{j-1}:\tau_j}|\bx_j,\boldtheta_j),
\]
where $\tau_0=0,\tau_{k+1}=n$, $\bx_j$ represents the part of the GMRF $\bx$ which belongs to the $j^{\mbox{\tiny th}}$ segment, and $\boldtheta = (\boldtheta_1^{\transpose},\dots,\boldtheta_{k+1}^{\transpose})^{\transpose}$ are the segment hyperparameters. Independent priors are taken on the members of $\Theta$ and the changepoints given their number. The prior taken on changepoints is assumed to have the product form 
\[
\pi_k^{\mbox{\tiny cp}}(\tau_1,\dots,\tau_k) = \prod_{j=0}^{k} \pi_k^{\mbox{\tiny cp}}(\tau_j|\tau_{j+1}).
\]
where $\tau_0=0,\tau_{k+1}=n$. Note that this prior is conditional on a given number of changepoints, $k$. The idea is to introduce a prior on $k$ and use the hierarchical form
\begin{equation}
\pi(k|\by) \propto \pi(\by|k) \pi(k) \label{eq:hierarchical_k}
\end{equation}
to find the most likely number of changepoints. Using this, the most likely positions for the changepoints can then be found.

\subsection{Recursively computing the posterior}

Let $L_j^{(k)} (t) = \Pr(\by_{t:n}|\tau_j=t-1, k)$. Then $L_j^{(k)}(t)$ is the probability of the data from time point $t$ onwards given the $j^{\mbox{\tiny th}}$ changepoint is at time $t-1$ and there are $k$ changepoints in total, meaning that there are $k-j$ changepoints between times $t$ and $n$. It is possible to compute $L_j^{(k)}(t)$ in a backward recursion;
\[
L_j^{(k)}(t) = \sum_{s=t}^{n-k+j}P(t,s)L_{j+1}^{(k)}(s+1)\pi_k^{\mbox{\tiny cp}}(\tau_j=t-1|\tau_{j+1}=s)
\]
with $j$ going from $k-1$ to $1$ and $t$ going from $n-k+j-1$ to $j+1$, where $P(t,s)=\pi(\by_{t:s})$ is the marginal likelihood of the segment $\by_{t:s}$. The marginal likelihood of $\by_{1:n}(=\by)$ under a $k$ changepoint model may be computed as
\begin{equation}
\Pr(\by_{1:n}|k) = \sum_{s=1}^{n} P(1,s)L_1^{(k)}(s+1)\pi_k^{\mbox{\tiny cp}}(\tau_1=s). \label{eq:marg_like_kchangepts}
\end{equation}

\subsection{Choice of changepoint prior and computational cost}

It will be necessary to compute $\pi(\by_{1:n}|k)$ for a range of values, say $k=0,\dots,K$ in order to do inference for $k$ using (\ref{eq:hierarchical_k}). This requires computational effort in $O(n^2 K^2)$ and storage requirements in $O(n K^2)$ which could be costly. Both of these may be reduced by choosing an appropriate changepoint prior. One such prior, as used and noted by~\citeasnoun{Fearnhead06}, is to take changepoint positions distributed as the even numbered order statistics of $2k+1$ uniform draws from the set $\{1,\dots,n-1\}$ without replacement. Doing this gives
\[
\pi_k^{\mbox{\tiny cp}}(\tau_1,\dots,\tau_k) =\frac{1}{Z_k}\prod_{j=0}^{k} \delta(\tau_j|\tau_{j+1})
\]
where $\delta(s|t)=t-s-1$ and the normalizing constant $Z_k=\binom{n-1}{2k+1}$. Using this prior restricts the dependence of the prior on the number of changepoints to the normalizing constant only, meaning that
\begin{eqnarray*}
L_{j+r}^{(k+r)}(t) &=& \sum_{s=t}^{n-\left[k+r-(j+r)\right]} P(t,s) L_{j+r+1}^{(k+r)}(s+1)\delta(\tau_{j+r}=t-1|\tau_{j+r+1}=s)\\
&=& \sum_{s=t}^{n-k+j} P(t,s)L_{j+r+1}^{(k+r)}(s+1)\times(s-t)\\
&=& \sum_{s=t}^{n-k+j} P(t,s)L_{j+1}^{(k)}(s+1)\times(s-t) =  L_{j}^{(k)} (t).
\end{eqnarray*}
Reusing these values gives a reduction by a factor of $K$ in computational effort and storage requirements. The recursions are now 
\begin{equation}
L_j^{(k)}(t) = \sum_{s=t}^{n-k+j} P(t,s)L_{j+1}^{(k)}(s+1)\delta(\tau_j=t-1|\tau_{j+1}=s) \label{eq:recursions_full}
\end{equation}
and 
\begin{equation}
\Pr(\by_{1:n}|k) = \sum_{s=1}^{n} P(1,s)L_1^{(k)}(s+1)\delta(\tau_0=0|\tau_1=s). \label{eq:recursions_prior_spec}
\end{equation}
Then (\ref{eq:recursions_prior_spec}) is divided by $Z_k$ to correctly normalize the prior and (\ref{eq:hierarchical_k}) is obtained by multiplying this by the prior weight for $k$ changepoints $\pi(k)$. This prior will be used in the examples later.

\subsection{Posterior of any changepoint}

Since the prior on changepoints makes the changepoint model factorizable, it is possible to write down the posterior distribution of $\tau_j$ conditional on $\tau_{j-1}$ and $k$;
\[
\Pr(\tau_j|\tau_{j-1},\by_{1:n},k) \propto P(\tau_{j-1}+1,\tau_j)L_{j}^{(k)}(\tau_j+1)\delta(\tau_{j-1}|\tau_j)/L_{j-1}^{(k)}(\tau_{j-1}+1).
\]
This is used for the forward simulation of changepoints once the backward recursions have been computed. It is also used to give the modal changepoint configuration as in the examples later.

\section{Approximate changepoint inference using INLAs}\label{sec:INLA_CP_RFRs}

 The essential ingredient of the approach presented in this paper is to replace the segment marginal likelihood $P(t,s)$ in the recursions
\[
L_j^{(k)}(t) = \sum_{s=t}^{n-k+j} P(t,s)L_{j+1}^{(k)}(s+1)\delta(\tau_j=t-1|\tau_{j+1}=s)
\]
with a segment marginal likelihood approximated using INLA. It is the case that $P(t,s)$ needs to be available in closed form to use a filtering recursions approach. This will never be the case for hierarchical GMRF models, which can account for within segment dependency. However, INLAs can be used to get a good approximation to $P(t,s)$ for hierarchical GMRF segment models. This opens up the opportunity for more realistic data models in many cases. There are also two other advantages: the posterior of the number of changepoints may be well approximated for model selection; and the posterior of any given changepoint can be computed to a high degree of accuracy. 

There are two potential drawbacks of the proposed approach however. The first is that it could require fitting a GMRF model to a very small amount of data, which could be limiting depending on the complexity of the within-regime model. For example, at least five data points would be required to make fitting a first order auto-regressive random field feasible. This means that for the approach to be reasonable it may be necessary to expect changepoints to be quite well separated. The second potential drawback contrasts with the first. For large amounts of data, using INLAs to compute the $n(n+1)/2$ segment marginal likelihoods necessary to compute the recursions (\ref{eq:recursions_full}) could be costly. The next section proposes a way to overcome both of these problems simultaneously, while still retaining almost all of the advantages of using a filtering recursions approach. This proposed solution is termed reduced filtering recursions for changepoints (RFRs).

\subsection{Reduced filtering recursions for changepoints} 

The main idea of RFRs is to compute the recursions at a reduced number of time points and approximate the full recursions (\ref{eq:recursions_full}). The recursion is not computed at every time point which takes $O(n^2)$ computation. The motivation is that if segments have a reasonable duration, changepoints can be detected in the region where they have occurred. 

An analysis using RFRs permits a changepoint to occur at some point in the reduced time index set $\{t_1,\dots,t_N\}$ with $t_i<t_j$ for all $i<j$. The assumption is that if there is a changepoint between $t_i$ and $t_{i+2}$ it can be detected at $t_{i+1}$. For convenience, define $t_0=0$ and $t_{N+1}=n$. The spacing of the $t_i$ is an important issue. If the spacing is too wide, then changepoints will not be detected. If the spacing is too narrow, many points are required to represent the entire data, thus increasing the computation time. If there is little prior knowledge of where changepoints occur the natural choice is equal spacing; $t_i=ig$ for some choice of $g$. The following example briefly explores the choice of $g$ and makes the preceding discussion clearer. 

\begin{figure}
\begin{center}
$
\begin{array}{cc}
\includegraphics[width=60mm]{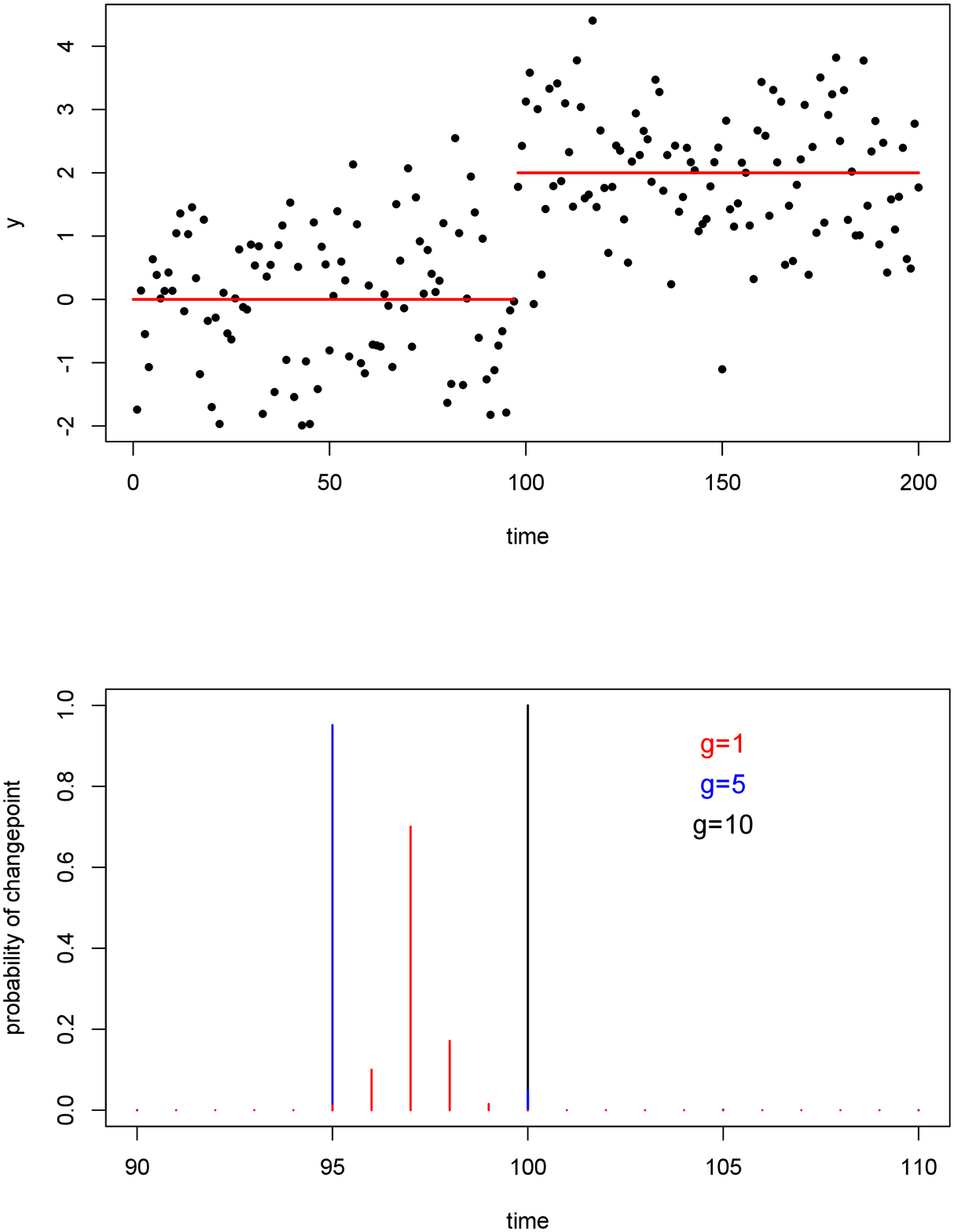} & \includegraphics[width=60mm]{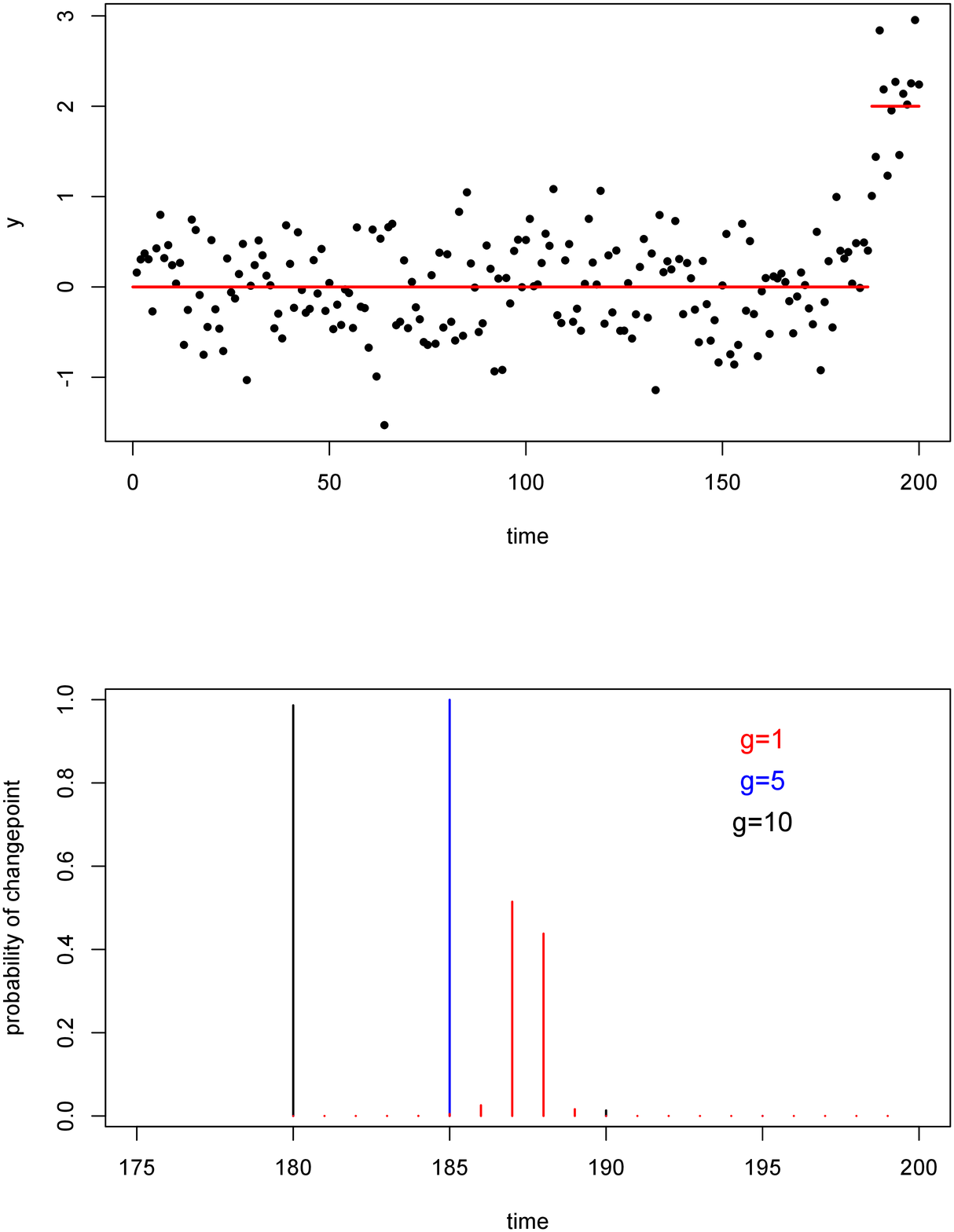}
\end{array}
$
\caption{Results when searching for one changepoint in simulated Gaussian data for $g=1,5,10$. It can be seen that the changepoint is detected at one of its closest neighbouring points in the reduced time index set.}\label{fig:PFR_example}
\end{center}
\end{figure}

Consider the data simulated from a Gaussian changepoint model shown at the top of Figure~\ref{fig:PFR_example}(a) with a clear change at 97. Searching for one changepoint, the bottom three plots in Figure~\ref{fig:PFR_example}(a) show the posterior probability of a changepoint for reduced time index sets given by $g=1,5,10$. Note that $g=1$ corresponds to the original recursions (\ref{eq:recursions_full}). For $g=5$ the changepoint is detected at 95 and $g=10$ detects it at 100. In both cases the changepoint is identified as the closest possible point to its actual position. Figure~\ref{fig:PFR_example}(b) shows a similar example, where this time one of the segments is very short (only 13 points). Again, the changepoint is identified at the closest possible position in the cases of $g=1,5$. In the case of $g=10$ it is the second closest, possibly due to the noise in the data contaminating the separation of the two regimes. The magnitude of the regime changes in these examples are large for illustration. If the magnitude of the change is small it will be necessary to have a longer segment to identify its presence with high power. It is also the case that if changes are short-lived, a large value of $g$ may cause changepoints to be missed.

\subsubsection{Recursions on the reduced time index set}

The changepoints are $\tau_1,\dots,\tau_k$. The reduced time index set is $\{t_1,\dots,t_N\}$. The changepoint prior is now defined on the set of numbers $\{1,\dots,N\}$ and we let $c_j=r$ if $\tau_j=t_r$. That is, $c_j$ corresponds to the changepoint position if time is indexed by $\{1,\dots,N\}$ whereas $\tau_j$ gives the changepoint position in the reduced time index set $\{t_1,\dots,t_N\}$.  Define
\[
R_j^{(k)}(r) = \Pr(\by_{t_r+1:n}|\tau_{j}=t_r, k).
\] 
For $r = N,\dots,k+1$
\[
R_k^{(k)}(r) = P(t_r+1,n) \delta(c_{k}=r|c_{k+1}=N+1).
\]
Then recursively, for $j=k-1,\dots,1$ and $r=N-k+j-1,\dots,j+1$
\[
R_j^{(k)}(r) = \sum_{s=r+1}^{N-k+j} P(t_r+1,t_s)R_{j+1}^{(k)}(s)  \delta(c_j=r|c_{j+1}=s).
\]
After computing these, the approximate marginal likelihood of the data conditional on $k$ changepoints follows as,
\[
\Pr(\by_{1:n}|k) \approx \sum_{s=1}^{N-k} P(1,t_s)R_1^{(k)}(s)\delta(c_0=0|c_1=s) / Z_k.
\]

When the grid spacing $g$ is not too large, that is $n$ is greater than a reasonable multiple of $g$, the approximation to the marginal probability of $k$ changepoints should be reasonable for the competing models. There are many computational savings with this approach. Using the RFRs decreases the number of marginal likelihood evaluations required to $n_r(n_r+1)/2$ where 
\[
n_r = \lfloor n/g + 1 - \mbox{I}(g=1) \rfloor.
\] 

\subsubsection{Distribution of any changepoint} \label{sec:INLA_distribution_any_changepoint}

When the maximum {\it a posteriori} number of changepoints has been found, it is determined where the changepoints are most likely to occur on the reduced time index set. The distribution of $c_j$ is
\begin{equation}
\Pr(c_j|c_{j-1},\by_{1:n},k) \propto P(t_{c_{j-1}}+1,t_{c_{j}})R_j^{(k)}(c_{j})\delta(c_{j}|c_{j+1})/R_{j-1}^{(k)}(c_{j-1}). \label{eq:pdist}
\end{equation}
Instead of generating samples of changepoints, our focus is to deterministically search for the most probable changepoint positions {\it a posteriori}. The first changepoint detected on the reduced time index set will be
\[
\hat{c}_1 = \arg\max_{c_1} \, \Pr(c_1|c_{0}=0,\by_{1:n},k).
\]
Conditioning on $\hat{c}_1$ the search proceeds for $c_2,\dots,c_k$ in the same way. In general,
\[
\hat{c}_j = \arg\max_{c_j} \, \Pr(c_j|\hat{c}_{j-1},\by_{1:n},k).
\]
This procedure is repeated until the $k$ changepoints $t_{\hat{c}_1},t_{\hat{c}_2},\dots,t_{\hat{c}_k}$ are found.

\subsubsection{Refining changepoint detection}\label{sec:refining}

After detecting changepoints on the reduced time index set, it is possible to refine the search and hone in on the most likely position of the changepoint. To begin, the changepoints obtained from the search above, $\tau_1^{(0)},\dots,\tau_k^{(0)}$ where $\tau_j^{(0)} = t_{\hat{c}_j}$, will all be multiples of $g$. Condition on the value of $\tau_2^{(0)}$ to update $\tau_1$. Compute
\[
P(1,\tau)P(\tau+1,\tau_2^{(0)})
\]
using INLAs for $\tau \in \{\tau_1^{(0)}-g+1,\dots,\tau_1^{(0)}+g-1\}$. Then take $\tau_1^{(1)}$ to be the $\tau$ which maximizes this. Similarly $\tau = \tau_j^{(1)}$ maximizes
\[
P(\tau_{j-1}^{(1)},\tau)P(\tau+1,\tau_{j+1}^{(0)}).
\]
This procedure can be carried out just once, or repeated until there is no difference between updates. 

This step does of course require additional computation. It may not be necessary in all cases to carry out a refined search. For example, the case of large $n$ and small $g$ would mean that refining the search would probably give little additional information. This approach should give near the global MAP for the changepoint positions. To ensure the global MAP is found it would be necessary to use some sort of Viterbi algorithm which would also use the approximated marginal likelihood values. 

\subsubsection{Simulation of changepoint positions}

The approximate methods discussed here are entirely simulation free. It may be useful to allow for simulation of changepoints from their joint posterior, once all of the marginal likelihoods have been approximated as in~\citeasnoun{Fearnhead06}. Introduction of the RFRs makes this a little more difficult here. We expect that for larger values of $g$ the distribution of the changepoints on the reduced time index set will be quite degenerate. This is since for a changepoint in a given region, it will always be detected at the same point in the reduced time index set. However, for smaller values of $g$ it may still be possible to simulate changepoints on the reduced time index set and refine their positions by simulating from a distribution which conditions on the two neighbouring changepoints- a stochastic version of the approach to refine the position of the changepoint discussed above (Section~\ref{sec:refining}).

\subsubsection{Exploring approximation error and computational savings in a DNA segmentation example} \label{sec:INLA_DNA_data}

To get a rough idea of the approximation error and the possible computational savings to be made by using RFRs, the methods were applied in a DNA segmentation task with a conditional independence model. This deviates from the general theme of the paper (to fit models relaxing conditional independence), however, it is included to offer some insight into RFRs in general. 

DNA sequence data is a string of the letters A,C,G and T representing the four nucleic acids, adenine, cytosine, guanine and thymine. Interest focuses on segmenting the sequence into contiguous segments characterized by their C$+$G content. It is assumed that within a segment the frequency of constituent acids follows a multinomial distribution, so that
\[
\pi(\by_{t:s}|\boldtheta) = \prod_{i=t}^s \theta_{\mbox{\tiny A}}^{\mbox{\tiny I}(y_i=\mbox{\tiny A})}\theta_{\mbox{\tiny C}}^{\mbox{\tiny I}(y_i=\mbox{\tiny C})}\theta_{\mbox{\tiny G}}^{\mbox{\tiny I}(y_i=\mbox{\tiny G})}\theta_{\mbox{\tiny T}}^{\mbox{\tiny I}(y_i=\mbox{\tiny T})}.
\]
With a $\mbox{Dirichlet}(\alpha,\alpha,\alpha,\alpha)$ prior on $\boldtheta_{(t:s)}$ the marginal likelihood for a segment is
\[
P(t,s) = \frac{\Gamma\{4\alpha\}}{\Gamma\{\alpha\}^4 \Gamma\{s-t+1+4\alpha\}} \prod_{j \in \{\mbox{\tiny A,C,G,T}\}} \Gamma\left\{n_j^{(t:s)} +\alpha\right\}
\]
where $n_j^{(t:s)}$ is the number of occurences of acid $j \in \{\mbox{A,C,G,T}\}$ in the segment from $t$ to $s$ inclusive.

\begin{figure}
\begin{center}
\includegraphics[width=80mm]{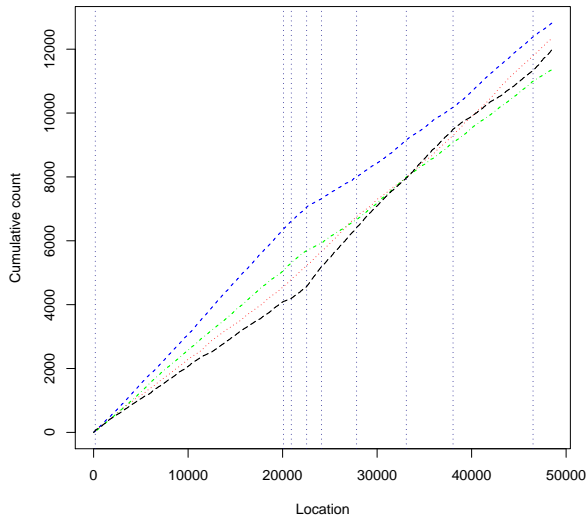}
\end{center}
\caption{Cumulative counts of A,C,G,T for the DNA data. Identified changepoints are overlain (vertical lines).} \label{fig:INLA_CP_lambda}
\end{figure}

The data analyzed is the genome of a parasite of the intestinal bacterium {\it Escherichia coli}. The sequence consits of 48,502 base pairs, and so will provide a good measure of the computational savings to be made for larger datasets when using RFRs. This data has previously been analyzed by~\citeasnoun{Boys04}, who implemented a hidden Markov model using RJMCMC to select the Markov order. Here however, a changepoint model assuming data in segments are independent is applied. Cumulative counts of the nucleic acids over location along the genome are shown in Figure~\ref{fig:INLA_CP_lambda}.

\begin{table}
\begin{center}
\begin{tabular}{l|cccccc}
\hline
\hline
 $g$ 		& 1 		& 5		 & 10			& 15		& 20		& 25\\
 \hline
Time taken (s) &1353.69 & 51.78 & 13.05 & 5.99 & 4.09 & 3.03\\
\hline
 Changepoints
 & $176$ 		& 176 		&176 		&176 		&176 		&176\\
 & 20092		& 20101		& 20092		& 20092		& 20092		&20092\\
 & 20920		& 20920		& 20920		& 20920		& 20920		&20920\\
 & 22546		& 22585		& 22546		& 22546		& 22546		&22546\\
 & 24119		& 24119		&24119 		& 24119		&24119 		&24119\\
 & 27831		&27831		 & 27831		& 27831		& 27831		&27831\\
 & $31226$ 	& 31226		&31226 		& 31226		& 31226		&$-$\\
 & 33101		&33101 		& 33101		& 33101		&33101 		&33089\\
 & 38036		& 38036		& 38049		& 38011		& 38036		&38036\\
 & 46536		&46536 		&46536 		& 46536		& 46501   		 &46501 \\
 \hline
 \hline
\end{tabular}
\end{center}
\caption{Location of changepoints and computing time for DNA segementation example. As $g$ increases there is little deviation in changepoint estimates. Reported changepoints are found after a refined search.} \label{tab:INLA_DNA_seg}
\end{table}

The RFRs were applied to this data using an equally spaced reduced time index set with $g= 1,5,10,15,20,25$. The prior taken on the number of changes was uniform on $\{ 0,1,\dots,20\}$. All runs were on a 2.66GHz processor written in C and the segment marginal likelihoods calculated in a step before the recursions were computed. Table~\ref{tab:INLA_DNA_seg} gives the identified changepoints and the computing time for each analysis. The value $g=1$ corresponds to filtering recursions on the entire data.  It can be seen that using RFRs does not appear to have a considerable effect on the detected changepoints. However, there are drastic differences in computing time- the RFRs for $g=25$ give a 450 fold decrease in computing time with respect to recursions on the full data set. It can be seen that as the value of $g$ increases there are slight deviations in the result. For example, with $g=25$, nine changepoints is most probable {\it a posteriori}, compared with ten for all other $g$ values. Also, for $g=5$ the second and  fourth changepoints are detected in a different position to all other values of $g$. This could be overcome by allowing for a wider search than just $g$ points either side in the refined search (Section~\ref{sec:refining}). Despite this, the computational savings are large, and the approach appears to successfully isolate the regions where changepoints occur in a large search space. 

It should be noted that the computation of the marginal likelihoods can be nested, although this was not done here. For example, the marginal likelihood calculations for $g=5$ could be reused for $g=10,15,\dots$ and likewise, some of the calculations for $g=10$ can be used for $g=5$ if it is desired to perform analysis for different values of $g$.

\subsection{Other approaches to save on computation}

The RFRs approach reduces the computation necessary to perform analysis for changepoint models by introducing an approximation to full filtering recursions. This is complimentary with another approach to reduce computation suggested by~\citeasnoun{Fearnhead06}. There, recursions are ``pruned'' by truncating the calculation of the backward recursion when the terms to be truncated will only contribute negligibly to the overall value. This occurs in situations where it is clear that a changepoint will have occurred before a certain time in the future, and so considering times after this point does not lead to gains in information. These ideas could be used together with the approach presented here to gain extra speed up in computation. In particular, RFRs are useful for situations where segment sizes are large while pruning ideas are useful when the number of changepoints is large (so that calculations may be truncated often), so combining the two approachess should be useful for large scale problems with regular changes and will lead to even greater computational savings than just using RFRs alone.

\section{Coal mining disasters}\label{sec:INLA_CP_coal_mining}

This data records the dates of serious coal-mining disasters between 1851 and 1962 and is a benchmark dataset for new changepoint approaches. It has been analyzed in~\citeasnoun{Fearnhead06},~\citeasnoun{Yang01},~\citeasnoun{Chib98},~\citeasnoun{Green95},~\citeasnoun{Carlin92} and~\citeasnoun{Raftery86}, amongst others. In all of these analyses it is assumed that observations arise from a Poisson process. This Poisson process is assumed to have intensity which follows a step function with a known or unknown number of steps. These steps or ``jumps'' in intensity occur at the changepoints. Other models have also been fit to this data. For example, a smoothly changing log-linear function for the intensity of the Poisson process:
\[
\lambda(t) = \nu \exp \{-\gamma t\}
\]
(see for example~\citeasnoun{Cox66} and the original source of this data~\citeasnoun{Jarrett79}). The log-linear intensity model would favour more gradual change, rather than the abrupt changes implied by changepoint models. There is an argument for some of the elements of such a model that allows for gradual change. Although, as noted in~\citeasnoun{Raftery86}, abrupt changes in this data are most likely due to changes in the coal mining industry at the time, such as trade unionization, the possibility of more subtle changes in rate could and should be entertained. A GMRF model applied to this data should be able to model gradual as well as abrupt change.


As in~\citeasnoun{Fearnhead06} a week is the basic time unit. The data spans 5,853 weeks over 112 years. The latent field is taken as AR(1). This allows for an inhomogeneous Poisson process within segments, opening up the possibility for gradual change. The rate of the Poisson process is related to the field through a log-link function. More specifically,
\[
y_i \sim \mbox{Poisson}(\lambda_i)
\]
where 
\[
\lambda_i = \exp\{ \alpha + x_i \},\quad i=1,\dots,n.
\]
The parameter $\alpha$ is an intercept and $x_i$ follows an AR(1) process with persistence parameter $\phi$.



Priors were chosen to loosely mimic the behaviour of the data. The priors chosen were
\begin{eqnarray*}
\sigma_{\bx}^{-2} &\sim& \mbox{Gamma}(4,0.01)\\
\kappa & \sim & \mbox{N}(3,1.89^2)\\
\alpha &\sim& \mbox{N}(0,10^2).
\end{eqnarray*}
where we have reparametrized $\kappa = \mbox{logit}\left(\frac{1+\phi}{2}\right)$. Following~\citeasnoun{Fearnhead06} and~\citeasnoun{Green95}, the prior on the number of changepoints was taken to be Poisson with mean 3.





\begin{figure}
\begin{center}
$
\begin{array}{cc}
\includegraphics[width=60mm]{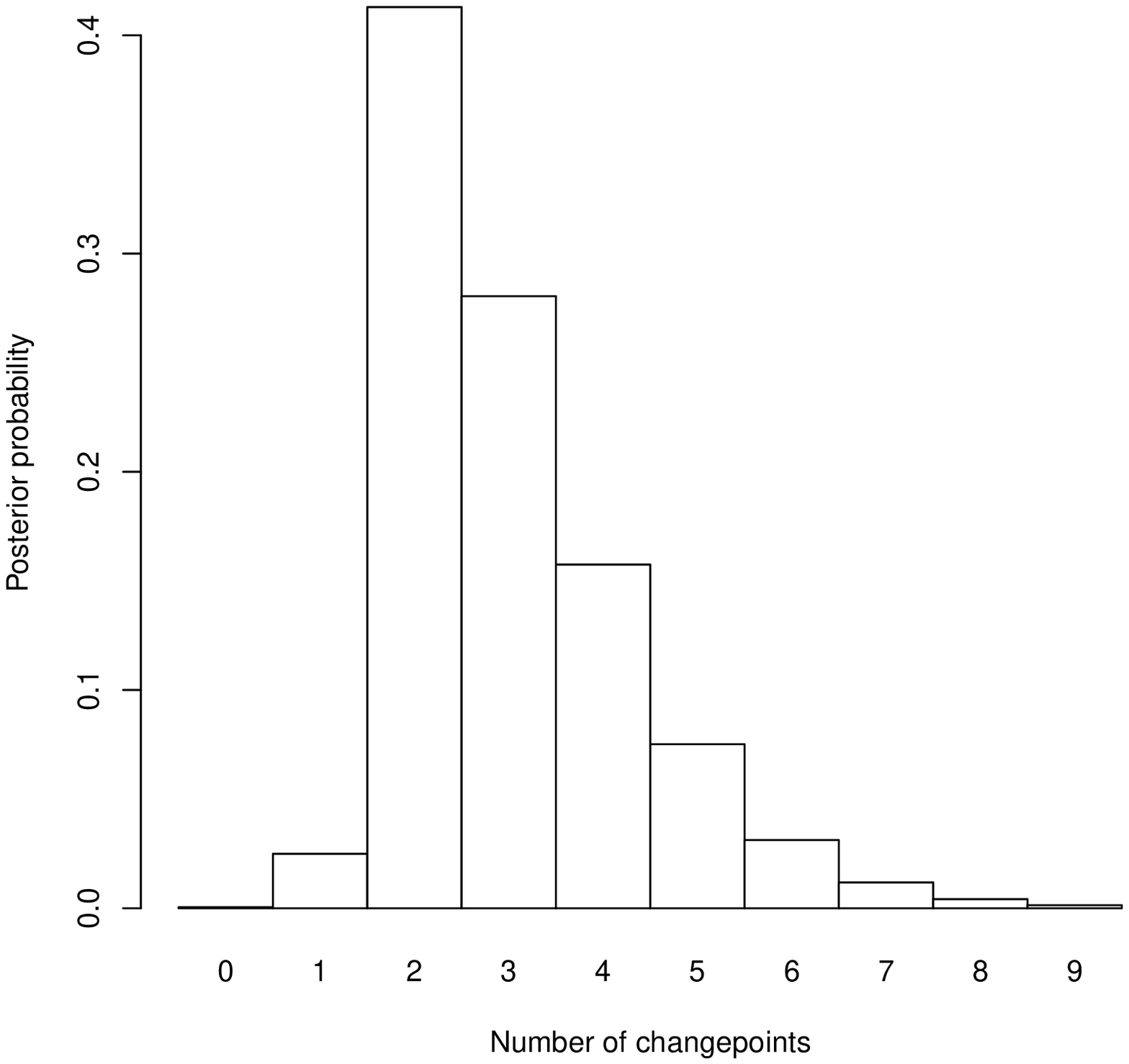} &
\includegraphics[width=60mm]{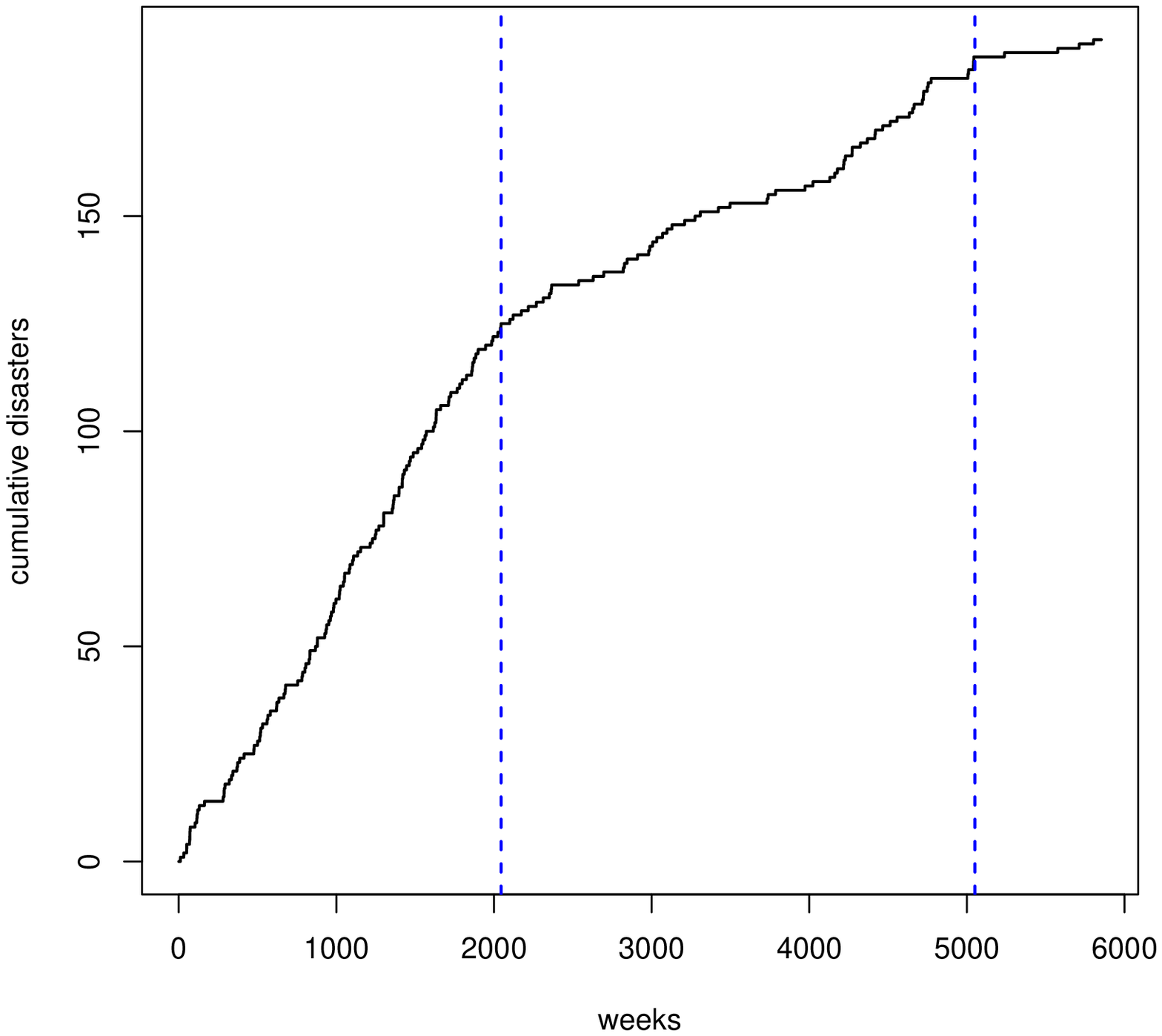}
\end{array}
$
\end{center}
\caption{Coal mining data: results from an anlysis using INLAs and $g=50$. The figure on the left shows the posterior distribution of the number changes while that on the right shows the cumulative counts of disasters and the changepoints indicated (blue dashed line).}\label{fig:INLA_coal}
\end{figure}

\begin{figure}
\begin{center}
\includegraphics[width = 60mm]{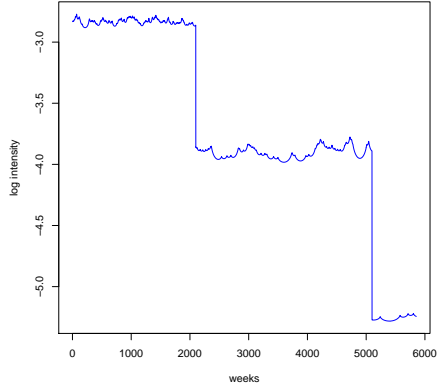}
\end{center}
\caption{Coal mining data: Inferred log intensity by week.} \label{fig:INLA_intensity}
\end{figure}

A spacing of $g=50$ was used. Figure~\ref{fig:INLA_coal} (a) shows the posterior distribution of the number of changepoints for the AR(1) latent field model. A two changepoint model is most likely, {\it a posteriori}. Figure~\ref{fig:INLA_coal} (b) shows the most likely position of these changepoints computed using the methods of Section~\ref{sec:INLA_distribution_any_changepoint}. A plot of the log intensity of the Poisson process over the entire 5,853 weeks is shown in Figure~\ref{fig:INLA_intensity}, obtained by conditioning on the MAP changepoint positions from the two changepoint model. From this it can be argued that a model accounting for gradual changes in the rate of disasters is not entirely unjustified. There appears to be small fluctuations of rate around a mean rate. These fluctuations are treated differently to the two abrupt changes that are detected by the GMRF model.

\begin{figure}
\begin{center}
$
\begin{array}{ccc}
\includegraphics[width=40mm]{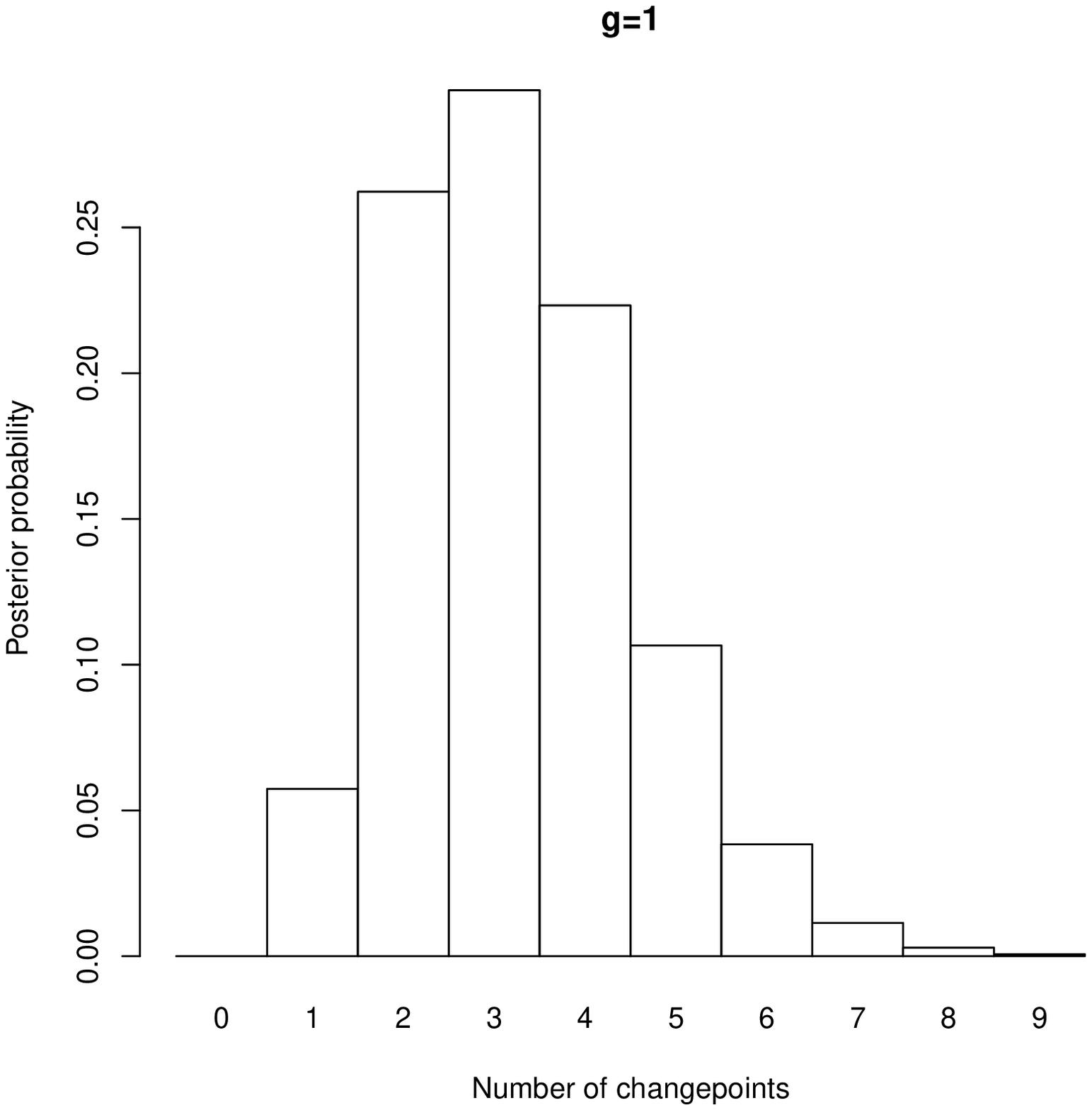} & 
\includegraphics[width=40mm]{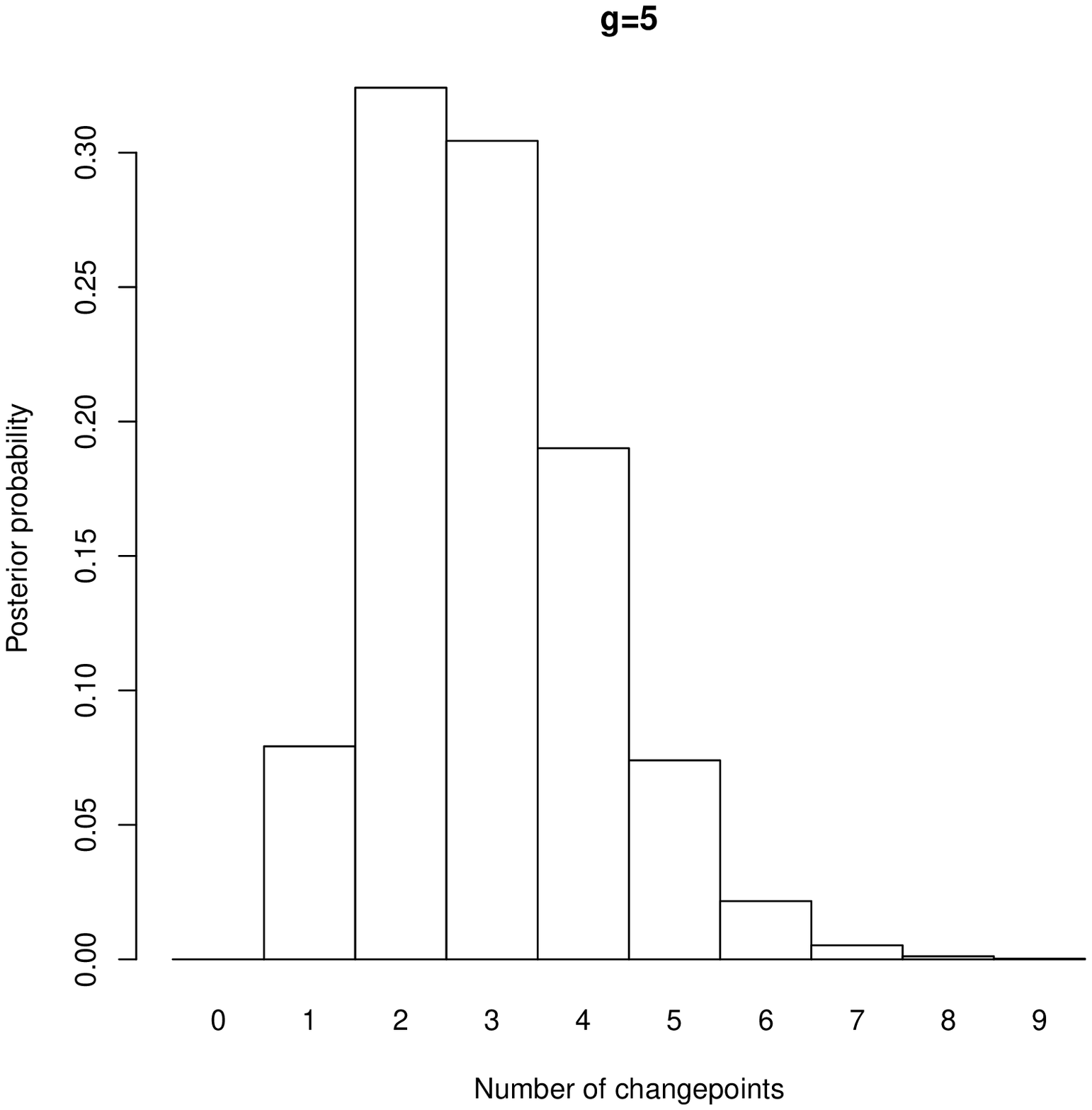}&
\includegraphics[width=40mm]{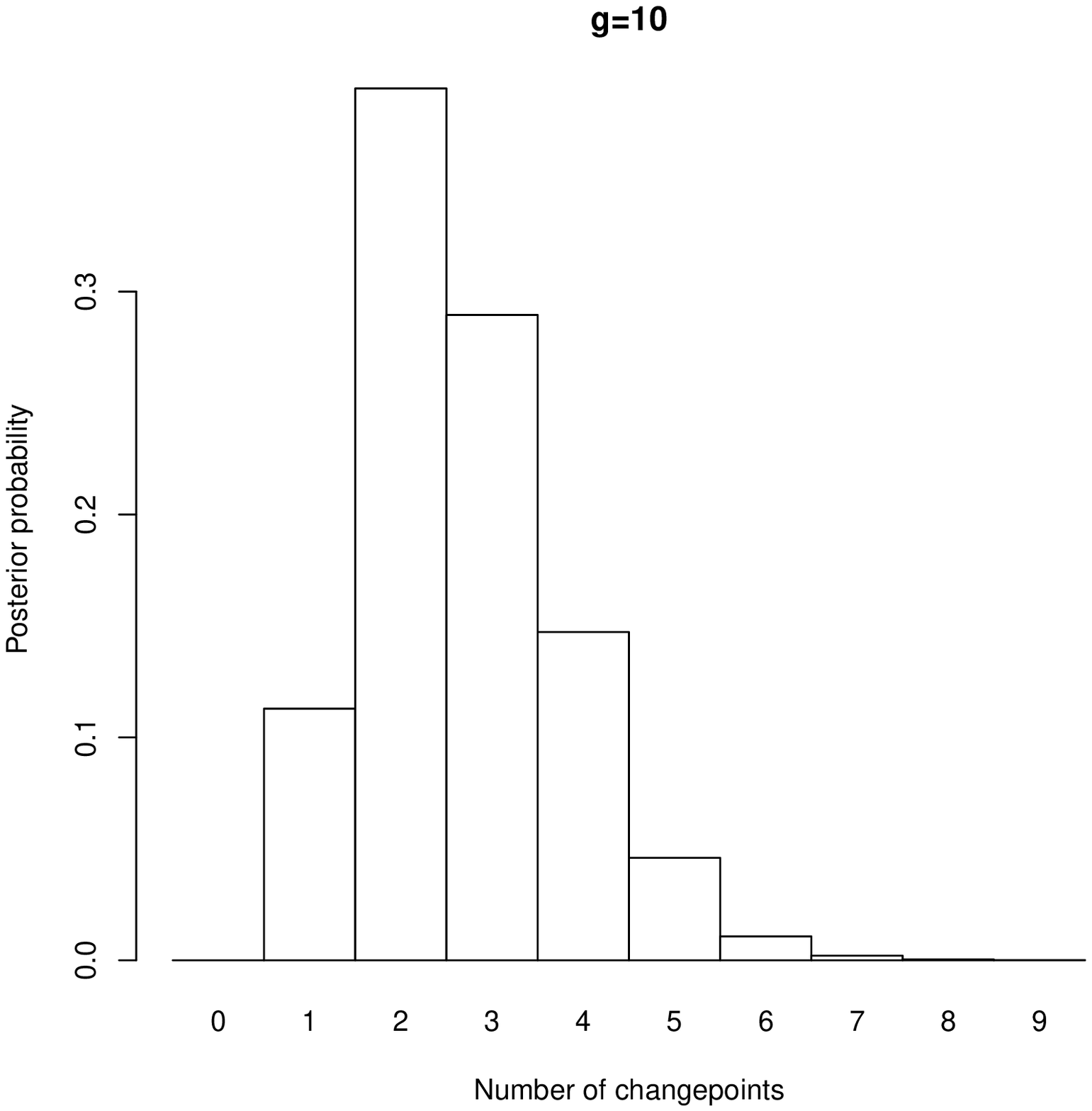} \\
\includegraphics[width=40mm]{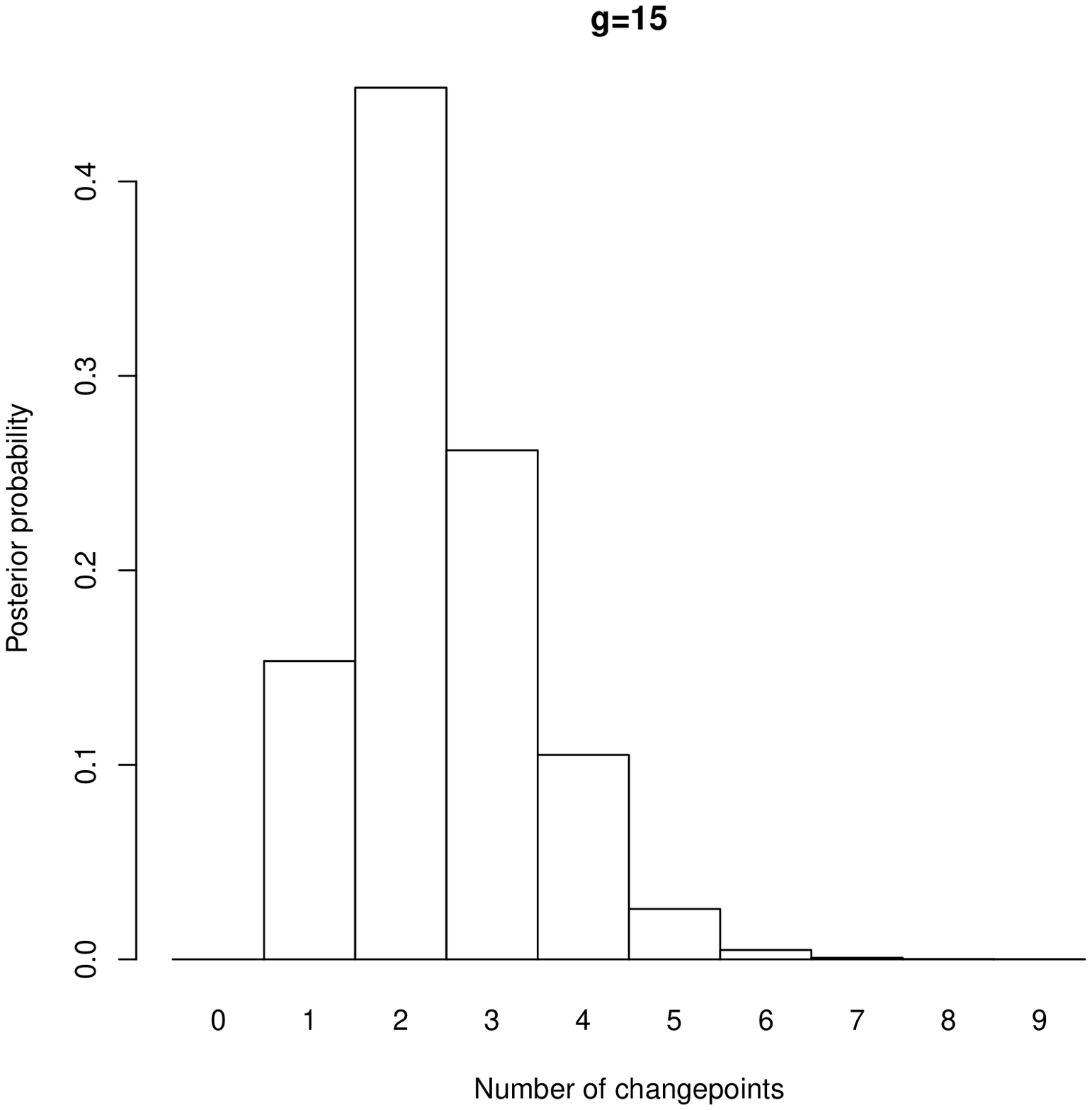} &
\includegraphics[width=40mm]{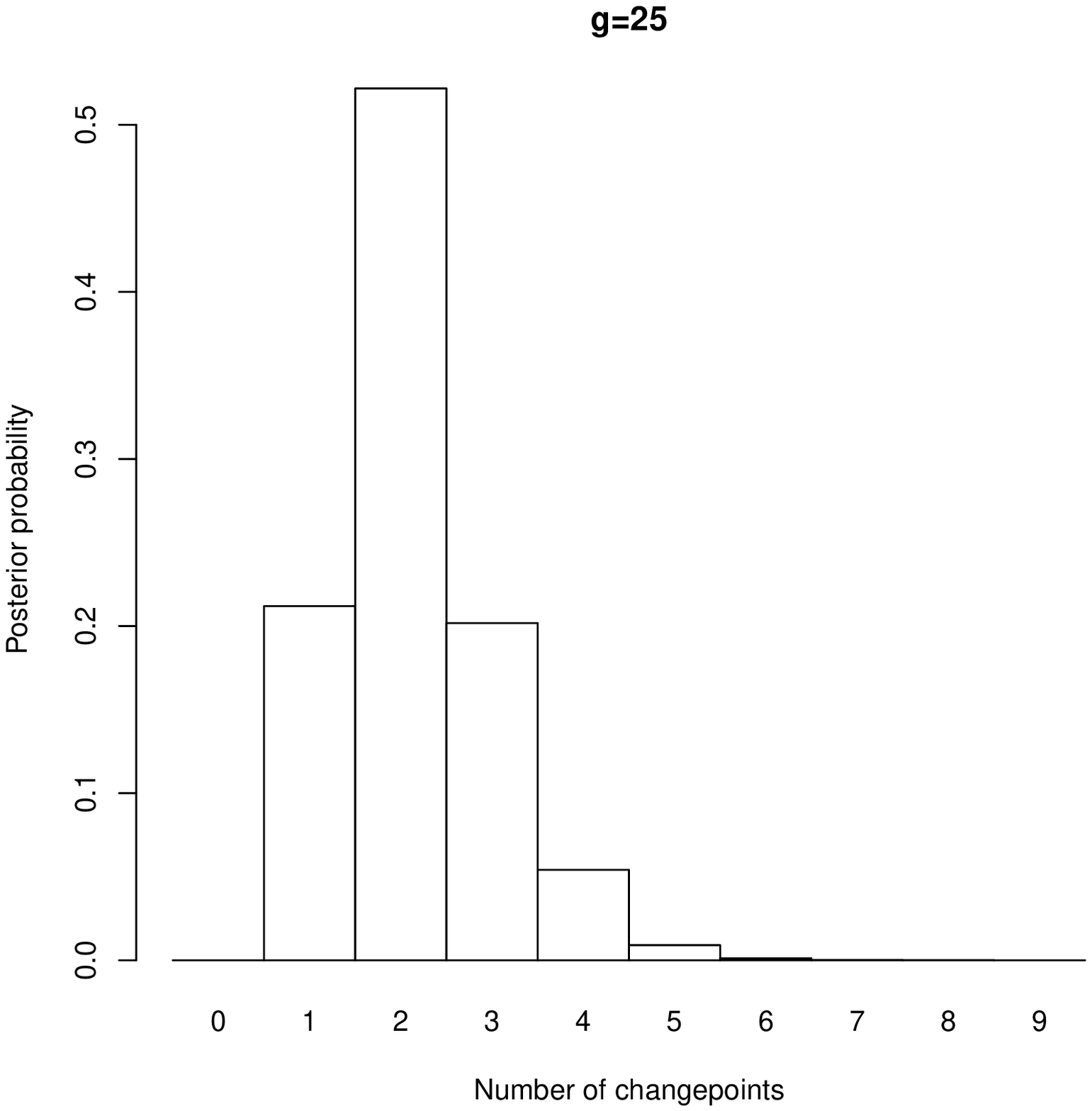} &
\includegraphics[width=40mm]{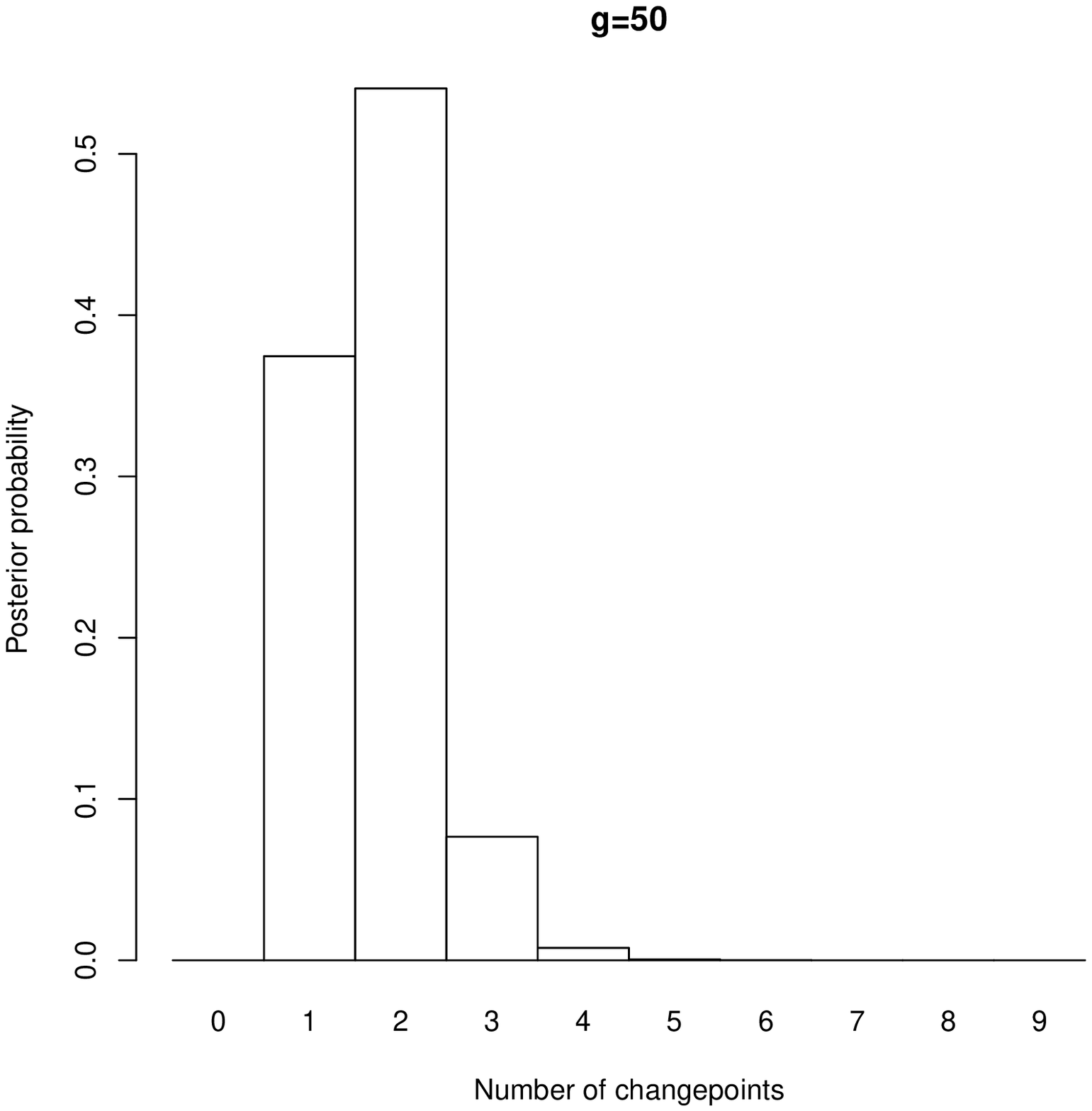}
\end{array}
$
\end{center}
\caption{Investigating approximation error in RFRs; results from analyses of coal mining disasters with different values of $g$ using the model from Fearnhead (2006).} \label{fig:INLA_RFR_approximation_error}
\end{figure}

There is a discrepancy between the posterior of the number of changepoints from RFRs given here and that given in~\citeasnoun{Fearnhead06} (see Figure 1(a) there) which both allowed changepoints at all possible points in the data. This is a good opportunity to further investigate the approximation error introduced by using RFRs. Figure~\ref{fig:INLA_RFR_approximation_error} shows the posterior number of changepoints obtained from using grids of size $g=1,5,10,15,25,50$ for the model and prior assumptions in~\citeasnoun{Fearnhead06}. It is clear that as the value of $g$ increases, the RFRs become less sensitive to small or short lived changes for this model, as might be expected. However, at large values of $g$ the ability to pick out two abrupt changes does not seem to diminish. As pointed out by one reviewer, a simple strategy for choosing $g$ is possible by exploiting the nesting ideas outlined in Section~\ref{sec:INLA_DNA_data}. Starting out with a large value of $g$ corresponding to a coarse search this may be gradually reduced to see how values of the approximated marginal likelihood for a given number of changepoints differs. Approximate marginal likelihoods computed for larger values of $g$ may be recycled in doing computations for the more refined searches. 

It is possible to compute approximate Bayes factors for the GMRF and independent data models conditional on there being a given number of changepoints. The marginal likelihood of the data conditional on $k$ changepoints is approximately
\[
\pi(\by_{1:n}|k) \approx \sum_{s=1}^{N-k} P(1,t_s)R_1^{(k)}(s)\delta(c_0=0|c_1=s) / Z_k.
\]
The different models are characterized by model assumptions and consequently the way in which the segment marginal likelihoods are computed;
\[
P_{\mbox{\tiny INLA}}(t,s) \quad \mbox{ and }\quad P_{\mbox{\tiny ANALYTIC}}(t,s).
\]
The approximate Bayes factor for the GMRF model versus the analytic model conditioning on $k$ changepoints is given by
\[
\mathcal{B}_k = \frac{\pi_{\mbox{\tiny INLA}}(\by|k)}{\pi_{\mbox{\tiny ANALYTIC}}(\by|k)}. 
\]
For a one changepoint model, this was $\mathcal{B}_1=4.63$ and for two changepoints it was $\mathcal{B}_2=5.25$. This implies that there is more support for the GMRF model in these cases, suggesting that modelling small scale variation in the rate of disasters is worthwhile. This supports the interpretation of Figure~\ref{fig:INLA_intensity}. It is well known that Bayes factors can be sensitive to prior assumptions. In this case prior hyperparameters have been chosen with care for both the independent data model and the GMRF model. A change in these choices could potentially lead to a different strength of conclusion as to which is the best model. However, it is still promising in this setting to see that modelling the dependency in the data appears worthwhile.

\section{Well-log data}\label{sec:INLA_CP_well_log_data}

\begin{figure}
\begin{center}
\includegraphics[width=100mm]{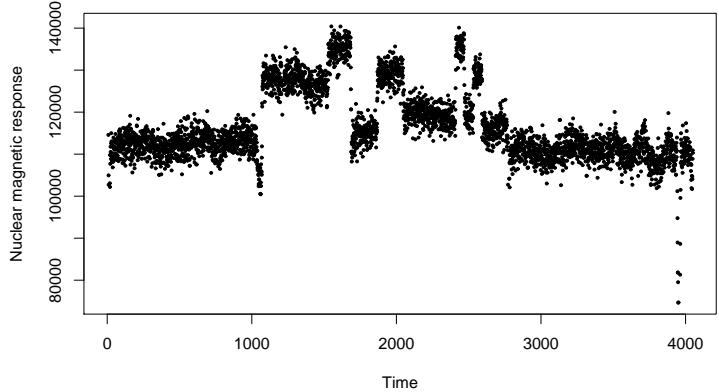}
\end{center}
\caption{Well-log data. Observations are the nuclear magnetic response recorded by a probe being lowered into a bore-hole in the Earth's surface.} \label{fig:well_log_data_plot}
\end{figure}

The Well-log data~\cite{ORuanaidh96} records 4050 measurements on the magnetic response of underground rocks obtained from a probe lowered into a bore-hole in the Earth's surface. The data is shown in Figure~\ref{fig:well_log_data_plot}. The model fitted here aims to account for dependency in the nuclear magnetic response as the probe is lowered into the bore-hole. This is an improvement on the independence model fitted in Section 4.2 of~\citeasnoun{Fearnhead06}; as the probe lowers, it moves through different rock strata and some will have greater depth than others. Therefore, it would be expected to see some correlation between observations arising from rock strata of the same type. Fitting this model can also reduce the detection of false signals as changepoints. See~\citeasnoun{Fearnhead03} for a discussion of the issue of outliers in Well-log data. 


Since this is a large data set $(n=4050)$ a larger value of $g$ should be used to isolate regions where changepoints occur. This vastly reduces the computational time required for the necessary approximations for data of this size. Analyses using $g=10,25,50$ were carried out, choosing the prior parameters using the information obtained from an analysis using MCMC and an independent data model. In each instance numerical instability prevented the recursions on the reduced time index set from being computed. This happened because the scale of the data is so large ($\sim 10^5$). In general, measures need to be introduced to prevent numerical instabilities in these types of recursions. In the computations of the RFRs a measure similar to those in~\citeasnoun{Fearnhead05} (changepoint models) and~\citeasnoun{Scott02} (hidden Markov models) was employed. This consisted of two steps to ensure stability. Firstly, compute
\begin{eqnarray*}
\frac{R_j^{(k)}(r)}{R_{j-1}^{(k-1)}(r+1)} &=& \sum_{s=r+1}^{N-k+j} \delta(c_j = r|c_{j+1} =s ) \exp\left\{\log P(t_r+1,t_s)+\log R_{j+1}^{(k)}(s) \right.\\
&& \qquad \qquad \qquad \qquad \qquad\qquad \qquad \qquad  \left. - \log R_{j-1}^{(k-1)}(r+1) \right\}
\end{eqnarray*}
and then
\[
\log R_j^{(k)}(r) = \log R_{j-1}^{(k-1)}(r+1) + \log \left(\frac{R_j^{(k)}(r)}{R_{j-1}^{(k-1)}(r+1)}\right).
\]
The reason these do not work here is that the large scale of the data means that $\log P(t_r+1,t_s)$ is much larger than usual, since it is the marginal likelihood of $g = 10,25,50$ points. It thus makes the argument to the exponential function in the first stabilizing equation cause instabilities at some points. This then carries through the remainder of the recursions. 

A simple way to overcome the issues is to just do an equivalent analysis of the data on a smaller scale, so that large $\log P(t_r+1,t_s)$ is avoided. Simply dividing the data by its sample standard deviation $s$ reduces the scale appropriately. The parameters for the prior specification were also adjusted to allow for the difference in scale to give the priors
\begin{eqnarray*}
\sigma_{\by}^{-2} & \sim & \mbox{Gamma}(1,0.01)\\
\sigma_{\bx}^{-2} & \sim & \mbox{Gamma}(1,0.01)\\
\kappa & \sim & \mbox{N}(5,(\sqrt{10})^2).
\end{eqnarray*} 
where $\kappa = \mbox{logit}\left(\frac{1+\phi}{2}\right)$. The prior on $\kappa$ here gives most prior weight to values of $\phi$ in $[0.9,1)$ (about 93\%). This will allow the possibility for the AR(1) GMRF model to closely approximate the behaviour of a random walk of order one. However, it still allows the freedom for the dependence pattern to vary across segments.~\citeasnoun{Fearnhead06} fits a random walk model of order one to this data, showing that a latent field can be robust to short lived changes and outliers for Well-log data. A uniform prior on $\{0,\dots,30 \}$ was taken for the number of changepoints.

\begin{figure}
\begin{center}
\includegraphics[width=60mm]{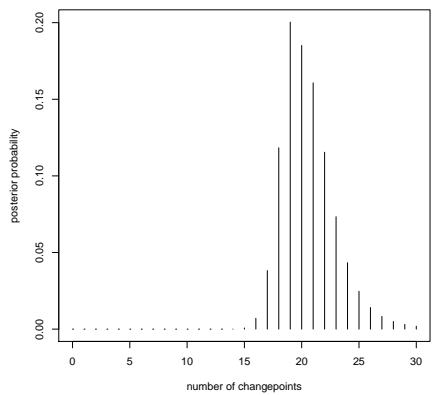}
\end{center}
\caption{Posterior of the number of changepoints for the Well-log data fitting an AR(1) GMRF model. This suggests the most likely number of changepoints {\it a posteriori} is 19.} \label{fig:well_log_posterior_k_GMRF}
\end{figure}

\begin{figure}
\begin{center}
$
\begin{array}{c}
\includegraphics[width = 140mm]{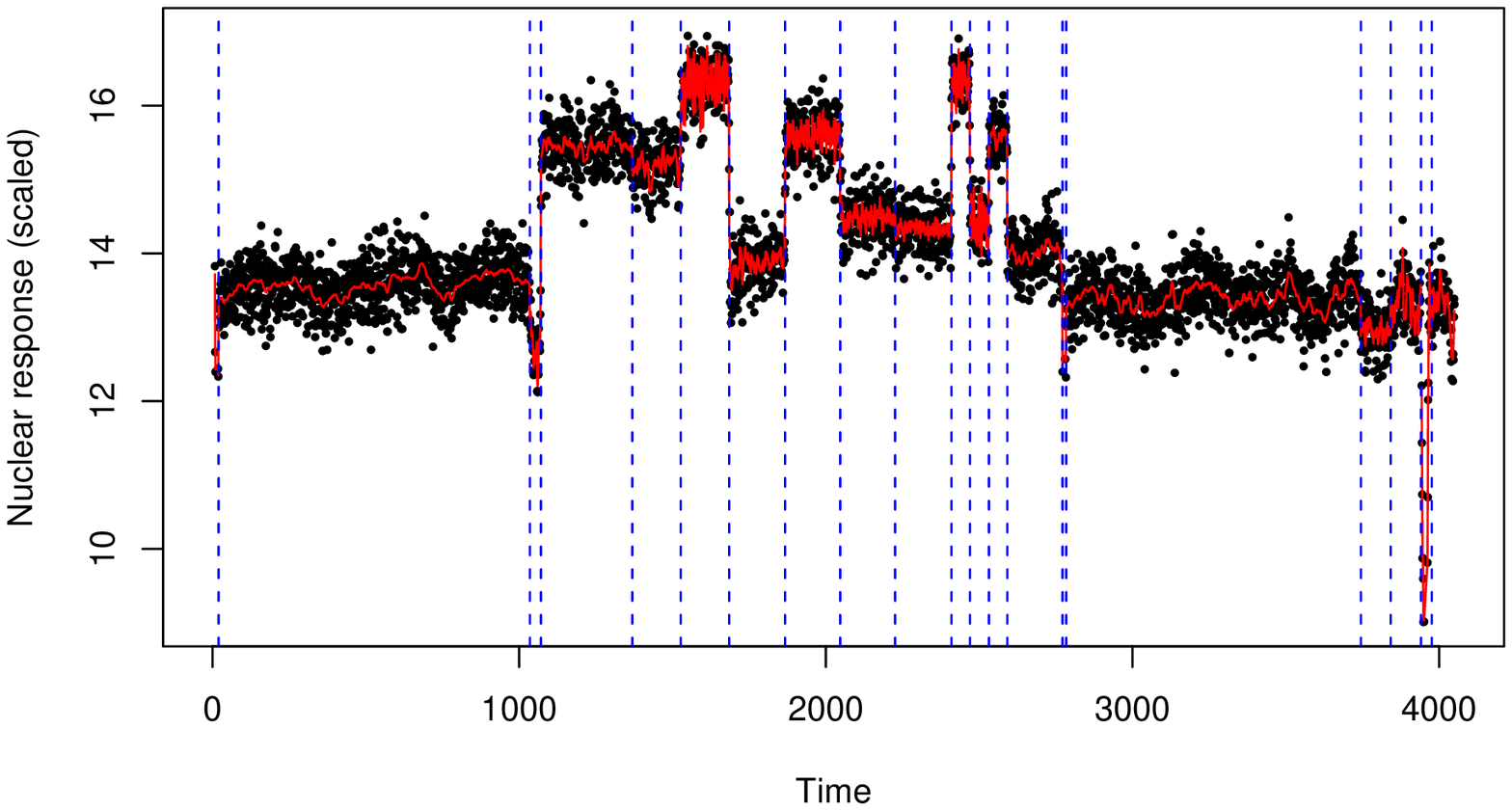}\\
\includegraphics[width = 140mm]{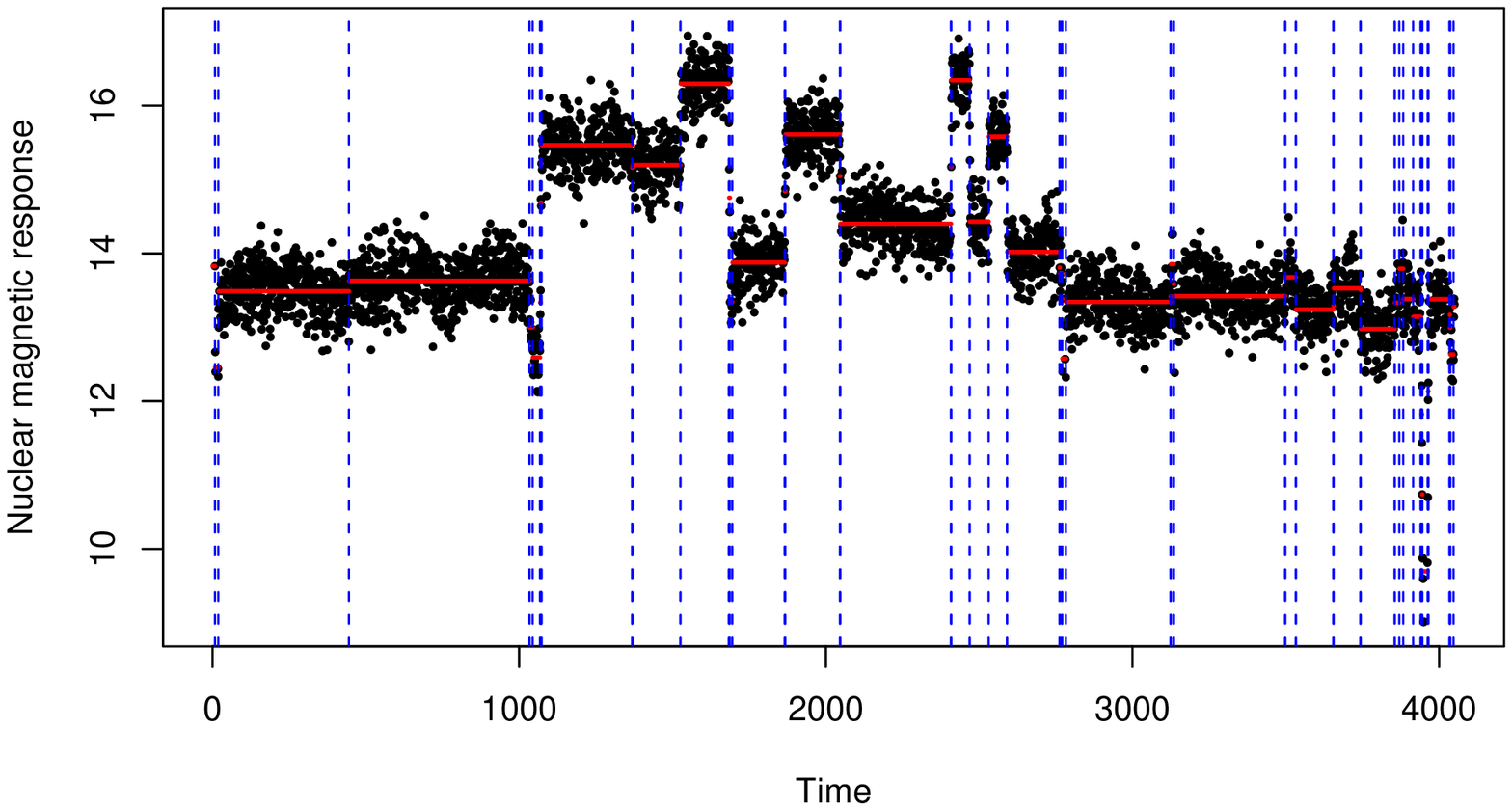}\\
\end{array}
$
\end{center}
\caption{Well-log data: results from RFRs and INLA (top) and independent data model.} \label{fig:well_log_GMRF_result}
\end{figure}


For the final analysis $g$ was taken to be 25. This reduced the necessary number of approximate marginal likelihood approximations from roughly $8.2\times 10^6$ (for $g=1$) to $1.3\times 10^4$; over 600 times less. The computations for these approximations took about a day of computing time. This appears lengthly, however this should be judged along with the fact that the model is more flexible and that the mean signal can be estimated at every point in the data. Figure~\ref{fig:well_log_posterior_k_GMRF} shows the posterior probability of the number of changepoints. The mode is at 19, but there appears to be support for up to 22. Conditioning on 19 changepoints, their locations were determined using the search strategy outlined in Section~\ref{sec:INLA_distribution_any_changepoint}. These locations were then refined to hone in on the actual changepoint positions. Conditioning on these positions inference was carried out for the latent field. This is shown in the top figure of Figure~\ref{fig:well_log_GMRF_result}. The field appears to follow the trend of the data closely, while the changepoint model caters for abrupt change.~\citeasnoun{Fearnhead06} compared the results of a first order random walk field to those from an independent Gaussian model for the data. Similarly, the results from the GMRF model here are compared with those obtained using an MCMC sampler with an independent data model on the Well-log data. For comparison, the 54 most likely changepoints (mode of posterior) were taken from the independent Gaussian model, and segment means were computed conditional on these (bottom of Figure~\ref{fig:well_log_GMRF_result}). It can be seen that the independent model is sensitive to changes in the mean and is conservative when inferring changepoints (more rather than less). The GMRF model however appears to be more robust to noisy data points and only infers changepoints when abrupt changes occur in the field.

\section{Stochastic volatility data} \label{sec:INLA_CP_stochastic_volatility}

The aim of this section is to explore whether INLAs and RFRs can be used to estimate changepoint models where segment observations are assumed to arise from a stochastic volatility model. To this end, the approach proposed could potentially used to detect shocks in financial and other time series. The segment model assumed is
\[
y_i \sim \mbox{N}\left(0,\beta^2 e^{x_i}\right), \quad i = 1,\dots,n,
\]
with $\bx$ following an AR(1) process with persistence parameter $\phi$ and innovation variance $\sigma_{\bx}^2$ where $2\log \beta$ may be interpreted as an intercept for the volatilities. Data in different segments are assumed independent, so that concern here is only in the complex intra segment correlation structure.

\begin{figure}
\begin{center}
\includegraphics[width = 150mm]{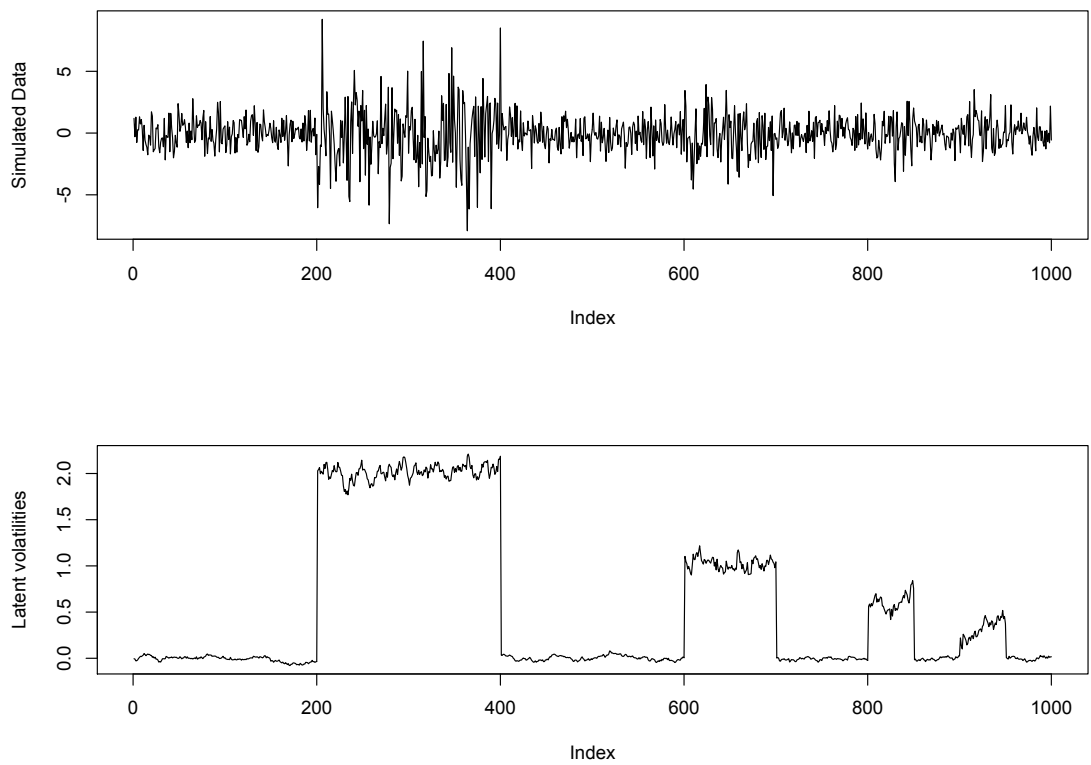}
\end{center}
\caption{Simulated stochastic volatility data with the corresponding latent log squared volatilities shown on the bottom.}\label{fig:StochVolRevision1}
\end{figure}

\begin{figure}
\begin{center}
\includegraphics[width = 100mm]{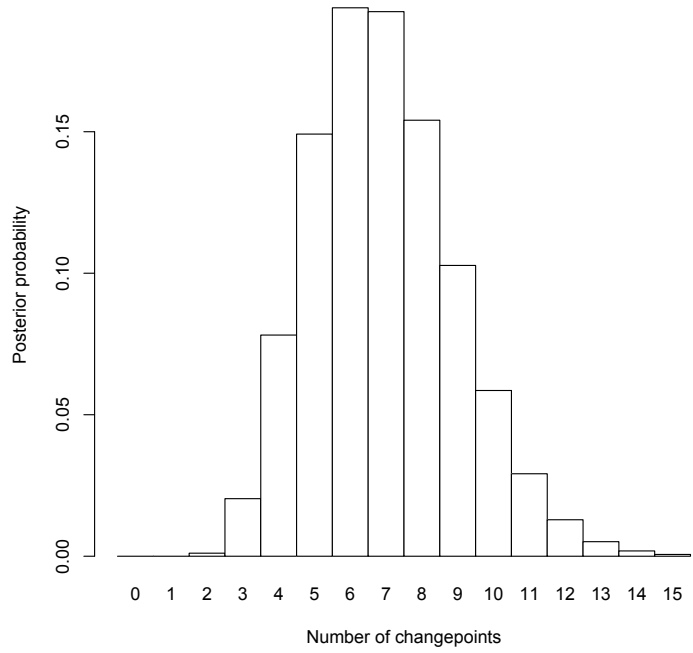} 
\end{center}
\caption{Posterior of the number of changepoints for simulated stochastic volatility data.}\label{fig:prior_sensitivity}
\end{figure}

\begin{figure}
\begin{center}
\includegraphics[width = 150mm]{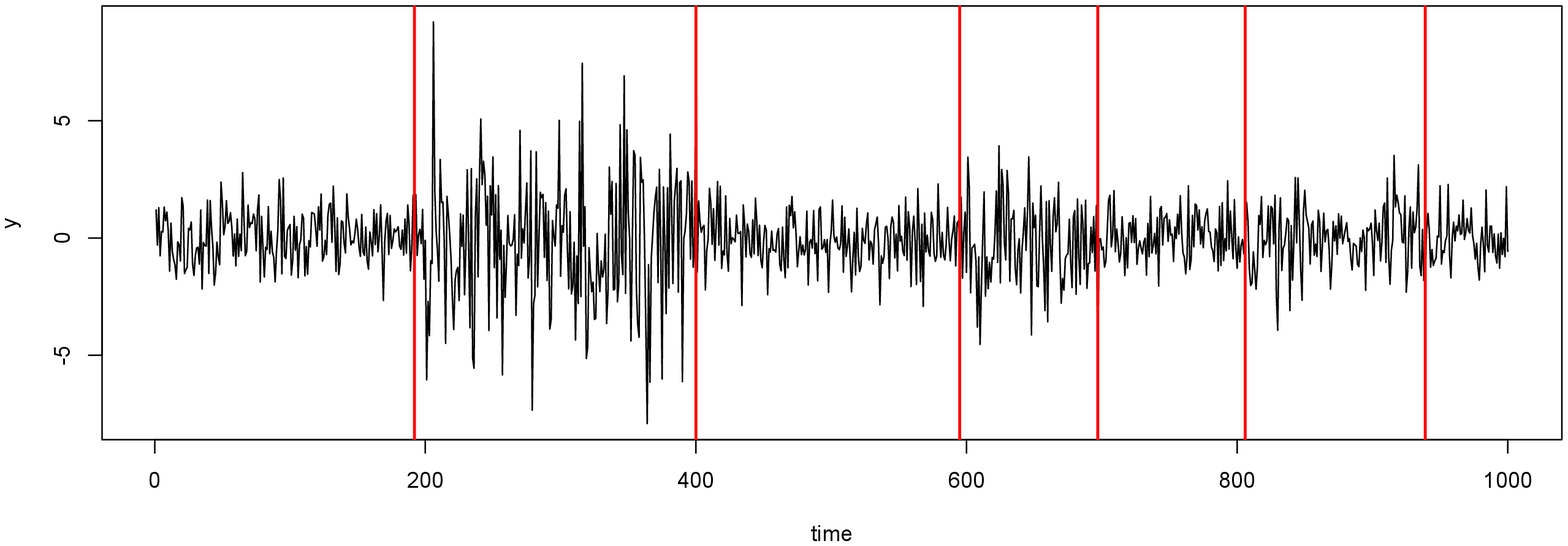}
\end{center}
\caption{MAP changepoint positions conditioning on six changepoints.}\label{fig:StochVolRevision2}
\end{figure}

The approach was applied to a simulated data set of length 1000 with eight changepoints at times 200, 400, 600, 700, 800, 850, 900 and 950, shown along with the corresponding log latent squared volatilities in Figure~\ref{fig:StochVolRevision1}. Segment parameters were chosen as outlined in Table~\ref{tab:svdata}. The length of the segments were reduced as well as the magnitude of the regime change to see how powerful the approach is in detecting smaller and smaller changepoints.

\begin{table}
\begin{center}
\begin{tabular}{c|ccccccccc}
\hline
\hline
 Segment & 1 & 2 & 3 & 4 & 5 & 6 & 7 & 8 & 9\\
 \hline
 $\phi$ & 0.9 & 0.8 & 0.9 & 0.7 & 0.8 & 0.9 & 0.8 & 0.9 & 0.8\\
 $2 \log \beta$ & 0 & 2 & 0 & 1 & 0 & 0.5 & 0 & 0.25 & 0  \\
  $\sigma_{\bx}$ & $0.01$ & $0.05$ & $0.01$ & $0.05$ & $0.01$ & $0.05$ & $0.01$ & $0.05$ & $0.01$\\
 \hline
 \hline
\end{tabular}
\end{center}
\caption{Segment parameters for simulated stochastic volatility data.} \label{tab:svdata}
\end{table}

The priors assumed for the analysis were
\begin{eqnarray*}
\sigma_{\bx}^{-2} & \sim & \mbox{Gamma}(30,0.02)\\
\kappa &\sim&\mbox{N}(3,1)
\end{eqnarray*}
and the computations were done for a reduced time index set with spacing $g=5$. Priors were chosen to loosely mimic the behaviour of the data. The prior chosen for the number of changepoints was $\mbox{Poisson}(5)$.


Figure~\ref{fig:prior_sensitivity} shows the posterior of the number of changepoints with most support for six changes, but reasonable support for any number from five to eight. The changepoint positions found using the search strategy of Section~\ref{sec:refining} while conditioning on six changepoints were 192, 400, 595, 697, 806 and 939. These changepoints are shown with the data in Figure~\ref{fig:StochVolRevision2}. The changepoints are not detected at their exact positions, but are roughly close to the true positions given the scale of the data. The two changepoints at 850 and 900 are missed by this estimation strategy. These two changes are small in magnitude and are barely noticeable by simply looking at the data. In this situation it may be too difficult for any estimation strategy to distinguish between noise and a weak signal.


\section{Discussion}

This paper demonstrates two new useful approximate methods for changepoint problems when the assumption of independent data is relaxed. The first of these was INLAs, a new approximate inference method for GMRFs due to~\citeasnoun{Rue09}. This allows the marginal likelihood for complex segment models to be evaluated approximately, so that it may be used for an approximate filtering recursions approach.

Some computational considerations led to the second proposed method. Instead of performing filtering recursions analysis on the entire data, RFRs were introduced so that recursions may be computed only on a reduced time index set, thus using all of the data, but only searching for changepoints in the general region where they occur. It was demonstrated that this method can be useful in cutting computation time for larger datasets by applying it to a DNA segmentation example with about 49,000 data points. 

The hybrid INLAs-RFRs methodology was applied to three different data examples. The first of these was an analysis of the coal mining disasters data where the model allowed for small scale variation in the intensity of the process and allowed for week to week dependency. This new model was more supported by the data than the usual step function intensity models which are often fitted. This was demonstrated by approximate calculation of Bayes factors for the GMRF model and the independent data model for one and two changepoint models. The GMRF model out-performed the independent data model in both cases. The second example was an analysis the Well-log data of~\citeasnoun{ORuanaidh96}. It was shown that allowing for segment dependency can be more robust to noisy observations, and that unnecessary changepoints (short lived changes, outliers $\mbox{etc.}$) are not inferred in this case. For the final example, the methods were applied to some simulated stochastic volatility data. Performance of the approach was promising in this case, however, there was difficulty in detecting some smaller changes.

It is worth noting again that RJMCMC would be practically infeasible for the data models considered here. This is since in addition to the issue of efficient sampling from hierarchical GMRF models (see~\citeasnoun{GMRFbook}), there is also segmentation of the data. Thus adding a new changepoint would require designing a reversible move between proposed and current field hyperparameters and in addition resampling field elements. Making moves of this type which exhibit good mixing would be challenging, and further diagnosing convergence would be difficult with the chains possibly requiring very long run times. This gives the approximate approach even more of an advantage. This is true especially in the case of models which require good corresponding proposal densities to perform well when it comes to MCMC, such as stochastic volatility models.


Overall, this paper has explored a promising new direction for estimation of changepoint models by creating a hybrid of two popular methods in their respective fields, namely INLAs in the GMRF field of study, and filtering recursions for sequential changepoint model estimation. Other data models are possible which have not been applied to any of the examples in this paper. For example, it is possible to have higher order Markov dependencies for random walk fields in the R-INLA package. Zero inflated Poisson and Binomial data models are also possible.

\section*{Acknowledgments}

The authors would like to thank the Editor, Associate Editor and two anonymous reviewers for their attentive reading of the manuscript and valuable suggestions which improved the presentation of the ideas in the paper. In particular we would like to thank the Editor for pointing out an inconsistency in results which led to us spotting a coding error in the DNA example. The first author would like to dedicate his work on this paper to the memory Cian Costello (1984-2010) who is enormously missed as a colleague and friend.

\bibliography{paper_bibliography,bibliography2,paper_bibliography_2,bibliography3,paper_bibliography_MCMC_cp}
\end{document}